\def\br{{\bf r}}
\def\brp{{\br '}}
\def\bR{{\bf R}}
\def\bRp{{\bR '}}
\def\bk{{\bf k}}
\def\bkp{{\bk '}}

\def\w0{\omega_0}
\def\k0{k_0}
\def\wk{\omega_k}
\def\wkp{\omega_{k'}}
\def\epkj{\epsilon_{\bk j}}
\def\epkjs{\epsilon_{\bk j}^\star}
\def\epkjp{\epsilon_{\bk 'j'}}

\def\ekj{\hat{e}_{\bk j}}

\def\ekjp{\hat{e}_{\bk 'j'}}

\def\akj{a_{\bk j}}
\def\akjd{a_{\bk j}^\dagger}

\def\des{\mid \uparrow 0_{\bk j} \rangle}
\def\desb{\langle \uparrow 0_{\bk j} \mid}

\documentclass[preprint,aps,showpacs]{revtex4}

\begin{document}

\title{Vacuum field correlations and three-body Casimir-Polder potential with one excited atom} 

\author{R. Passante\mbox{${\ }^{*}$}, F. Persico\mbox{${\ }^{**}$}, L. Rizzuto\mbox{${\ }^{**}$}}
\affiliation{
\mbox{${\ }^{*}$}Istituto di Biofisica - Sezione di Palermo, 
Consiglio Nazionale delle Ricerche, Via Ugo La Malfa 153, I-90146 Palermo, Italy \\
\mbox{${\ }^{**}$} INFM and Dipartimento di Scienze Fisiche ed Astronomiche, 
Universit\'{a} degli Studi di Palermo, Via Archirafi 36, I-90123 Palermo, Italy 
 }

\email{roberto.passante@pa.ibf.cnr.it}


\begin{abstract}
The three-body Casimir-Polder potential between one excited and two ground-state atoms is evaluated.
A physical model based on the dressed field correlations of vacuum fluctuations is used, generalizing a model
previously introduced for three ground-state atoms. Although the three-body potential with one excited atom is
already known in the literature, 
our model gives new insights on the nature of non-additive Casimir-Polder
forces with one or more excited atoms.
\end{abstract}

\maketitle

\section{\label{sec:1}Introduction}

The Casimir-Polder potential is a long-range interaction between neutral atoms or molecules in the vacuum,
arising from their common interaction with the electromagnetic radiation field \cite{CP48,P01,Spruch96}. 
Casimir-Polder forces are also
a manifestation of the existence of vacuum fluctuations \cite{CPP95}. An important property of these forces 
is that, in the case of three or more atoms/molecules, they are non-additive \cite{AZ60,PT85}.
The origin of the non-additive behaviour of the Casimir-Polder potential for 
ground state atoms can be attributed to the change
of the field zero-point energy \cite{PT94}, or to the existence of dressed zero-point field correlations \cite{CP97}
or to dressed field fluctuations around the atoms \cite{PP99}. Casimir-Polder forces have also
been obtained by semiclassical methods in the framework of stochastic electrodynamics \cite{Boyer69}.
Recently, there has been also interest in Casimir-Polder forces 
when one or more atoms are in an excited state, where contributions from resonant poles arise \cite{PT93,FSBD95}. 
When one atom is excited, a time-dependent Casimir-Polder potential is in general expected, 
even if it may take quite a long time before the potential settles to a stationary value \cite{RPP04}.
This paper deals with the three-body potential during this time in which the excited atom can be approximately
treated as if in a metastable state.

In this paper we introduce a physical model in order to calculate the three-body Casimir-Polder potential 
between one excited and two ground-state atoms, and in order to understand its origin. 
The method we use is based on the connection between spatial field correlations 
and the Casimir-Polder potential \cite{PT93a, PPR03}. We have already exploited a similar method in the
case of three ground-state atoms \cite{CP97}. However, when one or more atoms are excited
the situation changes due to the possibility of a resonant atom-field interaction. We shall show
that the model based on dressed field correlations, and the underlying physical interpretation
of many-body Casimir-Polder forces, applies with appropriate changes also to the
case of excited atoms.

We consider a system of three atoms in vacuo, labeled A, B, C, interacting with the electromagnetic
radiation field. We use the multipolar coupling scheme, which is more appropriate for dealing with this kind
of problems, because this coupling scheme includes electrostatic interactions in the transverse fields
 \cite{CPP95}. We take atoms A and B to be in their ground state and C in its excited state.
Rather than using conventional perturbation theory and starting from a three-atom Hamiltonian
to obtain the total energy shift as a function of the atomic configuration \cite{PT95}, we shall adopt here
a completely different approach in which this shift is related to the interaction between all possible pairs
of atoms, correlated by the vacuum fluctuations which are changed by the presence of the third atom.
We first concentrate on the excited atom C, approximated as a two-level atom; 
its interaction with the radiation field is described by
the Hamiltonian

\begin{eqnarray}
H &=& \hbar \w0 S_{z} +
\sum_{\bk j} \hbar \wk \akjd \akj + \nonumber
\\
&+& 
\sum_{\bk j} \left( 
\akj e^{i\bk \cdot \br_{C}}- \akjd e^{-i\bk \cdot \br_{C}} 
\right) 
\left( \epkj S_{+} - \epkjs S_{-} \right)
\label{eq:1}
\end{eqnarray}         
where $\w0 =c\k0$ is the transition frequency of the atom,
$\br_C$ is its position and
$S_{z}$, $S_{+}$ and $S_{-}$ are the pseudospin atomic 
operators.
The coupling constant $\epkj$, in the multipolar
coupling scheme is given by  

\begin{equation}
\epkj = -i \left( \frac{2\pi\hbar c k}{V} \right)^{1/2} 
\mbox{\boldmath $e$}_{{\bf k}j}\cdot \mbox{\boldmath $\mu$}^C
\label{eq:2}
\end{equation}
where $\mbox{\boldmath $\mu$}^C$ is the transition dipole
moment of atom C between the two levels considered
and $\ekj$ are the polarization unit vectors.

Due to the interaction with the field, the bare excited state 
$\des$ of atom C, whose spontaneous decay is assumed to be slow with
respect to other relevant processes, becomes dressed, and the second-order
non-normalized  dressed excited state is
\begin{eqnarray}
\des_D &=& \des + \frac 1\hbar \sum_{\bk j}
\frac {\epkjs e^{-i\bk \cdot \br_C}}{\w0 -\wk} 
\mid \downarrow \bk j \rangle
\nonumber \\
&+& \frac 1{\hbar^2} \sum_{\bk j \bkp j'}
\frac {\epkjs \epkjp e^{-i(\bk +\bkp) \cdot \br_C}}{(\w0 -\wk )(\wk +\wkp )}
\mid \uparrow \bk j \bkp j' \rangle
\label{eq:3}
\end{eqnarray}

We now evaluate the average value
of the spatial correlation of the Fourier components of the electric field
on the dressed state (\ref{eq:3}) 
\begin{eqnarray}
{\ }_D \desb E_\ell (\bk j, \br ) E_m (\bkp j', \brp ) \des_D 
= -\left( \frac {2\pi}V \right)^2 \left( \ekj \right)_\ell \left( \ekjp \right)_m
 \left( \ekj \right)_n \left( \ekjp \right)_p \mu^C_n \mu^C_p k k'
\nonumber \\
\times \left\{ \frac 1{k+k'} \left( \frac 1{\k0 -k} +\frac 1{\k0 -k'} \right)
e^{i\bk \cdot \bR} e^{i\bkp \cdot \bRp}
- \frac 1{(\k0 -k)(\k0 -k')} e^{i\bk \cdot \bR} e^{-i\bkp \cdot \bRp}
+ c.c. \right\}
\label{eq:4}
\end{eqnarray}
where $\bR = \br -\br_C$ and $\bRp = \brp -\br_C$. 
This function is similar to that obtained for a ground-state atom,
except for the presence of the pole at $\k0$.
Later we shall use
this expression for the calculation of the three-body Casimir-Polder
potential. After summation of (\ref{eq:4}) over the field modes
$(\bk j, \bkp j')$ and straightforward algebraic 
calculations involving polarization
sums and angular integrations, we obtain the spatial correlation function
of the electric field
\begin{eqnarray}
&\ &{\ }_D \desb E_\ell (\br ) E_m (\brp ) \des_D
= \sum_{\bk j \bkp j'} {\ }_D \desb E_\ell (\bk j, \br ) E_m (\bkp j', \brp ) \des_D 
\nonumber \\
&=& -\frac 1{\pi^2} \mu^C_n \mu^C_p F_{\ell n}^R F_{mp}^{R'} \frac 1{RR'}
\int_0^\infty \! dk \int_0^\infty \! dk' 
\left\{ \frac 1{\k0 -k} \left( \frac 1{k+k'} +\frac 1{k' -k} \right) \right.
\nonumber \\
&+& \left.  \frac 1{\k0 -k'} \left( \frac 1{k+k'} +\frac 1{k -k'} \right) \right\}
\sin kR \sin kR' + c.c.
\label{eq:5}
\end{eqnarray}
where we have defined the differential operator acting on $R$

\begin{equation}
F_{\ell n}^R = \left( -\delta_{\ell n} \nabla^2 +\nabla_\ell \nabla_n
\right)^R
\label{eq:6}
\end{equation}
We now integrate over $k'$ the first term inside the curly bracket and 
over $k$ the second term. After some algebra, we obtain
\begin{eqnarray}
&\ &{\ }_D \desb E_\ell (\br ) E_m (\brp ) \des_D
\nonumber \\
&=& \frac 2\pi \mu^C_n \mu^C_p F_{\ell n}^R F_{mp}^{R'} \frac 1{RR'}
\left( P\int_0^\infty \! dk \: \frac {\sin k(R+R')}{k-\k0}
+2\pi \sin \k0 R \sin \k0 R' \right)
\label{eq:7}
\end{eqnarray}
where the last term in (\ref{eq:7}) stems from the resonance at $\k0$
in the frequency integrations.
It is easy to show that (\ref{eq:7}) can be expressed in the following form

\begin{eqnarray}
{\ }_D \desb E_\ell (\br ) E_m (\brp ) \des_D &=& \frac {\hbar c}\pi
F_{\ell n}^R F_{mp}^{R'} \frac 1{RR'} \int_0^\infty \! du \: \alpha^C_{mp}(iu)
e^{-u(R+R')}
\nonumber \\
&+& 2 \mu^C_n \mu^C_p F_{\ell n}^R F_{mp}^{R'} 
\frac {\cos \k0 (R-R')}{RR'}
\label{eq:8}
\end{eqnarray}
where $\alpha^C_{mp}(iu)$ is the dynamical polarizability of atom C
in the excited state, extended to imaginary values of the frequency

\begin{equation}
\alpha^C_{mp}(iu) = \frac {-2\k0 \mu^C_m \mu^C_p}{\hbar c (\k0^2 +u^2)}
\label{eq:9}
\end{equation}

Equation (\ref{eq:8}) shows that the spatial correlation function consists of two terms.
The first term on the RHS  is of the same form as in the case of a ground-state atom 
(with the appropriate dynamic polarizability, which is that of the ground state rather
than (\ref{eq:9})), and includes contributions from all field modes. 
The second term stems from the resonance at $k=\k0$ and oscillates in space with 
frequency $c\k0$; no such a term is present in the case of a ground-state atom.

We now proceed to evaluate the three-body force, generalizing the model of \cite{CP97} 
to the present case in which one atom is excited. The physical model discussed in 
\cite{CP97} for three ground-state atoms describes
the three-body force as the classical interaction between
the instantaneous dipole moments within all possible pairs of the three
atoms; these dipole moments are induced and correlated by the (correlated) vacuum 
fluctuations dressed by the third atom. 
The connection between the induced dipole moment and the vacuum electric field is

\begin{equation}
\mu_i(\bk ) = \alpha (k) E_i(\bk j, \br )
\label{eq:9a}
\end{equation}
where the atom is supposed isotropic for simplicity, and $\alpha (k)$ is its dynamic
electric polarizability.
In this model the interaction between two dipoles (say A and B) is that of classical
dipoles oscillating at the frequency of the field modes that induce them \cite{CP97,CP96}
(see these references for more details on the model used).
This interaction is proportional to the well-known potential tensor \cite{BW80}
\begin{eqnarray}
&\ &V_{i\ell} (\bk ,\bkp ,\br =\br_B -\br_A ) 
\nonumber \\ 
&=& - \frac 12 \left( \left( \delta_{i\ell} -\hat{r}_i \hat{r}_\ell \right)
k^2\frac {\cos kr}r - 
\left( \delta_{i\ell} -3\hat{r}_i \hat{r}_\ell \right)
\left( \frac {k\sin kr}{r^2} + \frac {\cos kr}{r^3} \right)
+ (k \leftrightarrow k') \right)
\nonumber \\
&=& -\frac 12 \left( F_{i\ell}^r \frac {\cos kr}r
+ F_{i\ell}^r \frac {\cos k'r}r \right)
\label{eq:10}
\end{eqnarray}
In the case of three ground-state atoms, a final simmetrization over the three identical atoms
is performed in order to take in account the symmetric role of the three atoms.
 
In the case in which one atom (say C) is excited, the permutational symmetry is broken and
the presence in (\ref{eq:8}) of the extra term (oscillating
in space) suggests the appearance of an additional contribution to the correlation of the
dipole moments of the two other atoms (A and B). 
Only the resonant field mode with $k=\k0$ is effective for this
additional correlation between the induced dipole moments.

In order to evaluate the interaction potential in the asymmetric case with atom C excited,
we must go back to equation (\ref{eq:4}), that we can write in the following form,
in order to identify and separate resonant and nonresonant contributions

\begin{equation}
{\ }_D \desb E_\ell (\bk j, \br_A ) E_m (\bkp j', \br_B ) \des_D
= \langle E_\ell (\bk j, \br_A ) E_m (\bkp j', \br_B ) \rangle_{r}
+ \langle E_\ell (\bk j, \br_A ) E_m (\bkp j', \br_B ) \rangle_{nr}
\label{eq:11a}
\end{equation}
where
\begin{eqnarray}
&\ & \langle E_\ell (\bk j, \br_A ) E_m (\bkp j', \br_B ) \rangle_{r} 
= \left( \frac {2\pi}V \right)^2 \left( \ekj \right)_\ell \left( \ekjp \right)_m
 \left( \ekj \right)_n \left( \ekjp \right)_p \mu^C_n \mu^C_p k k'
\nonumber \\
&\times& \left\{ \frac 1{k+k'} \left( \frac 1{k - \k0} +\frac 1{\k0 +k}
+\frac 1{k'-\k0} +\frac 1{k'+\k0} \right)
e^{i\bk \cdot \bR_{AC}} e^{i\bkp \cdot \bR_{BC}} \right.
\nonumber \\
&+& \left. \left( \frac 1{(k -\k0 )(k'-\k0 )}+  \frac 1{(k +\k0 )(k'+\k0 )} \right)
e^{i\bk \cdot \bR_{AC}} e^{-i\bkp \cdot \bR_{BC}}
+ c.c. \right\}
\label{eq:11b}
\end{eqnarray}
is the {\it resonant} part, and
\begin{eqnarray}
&\ & \langle E_\ell (\bk j, \br_A ) E_m (\bkp j', \br_B ) \rangle_{nr} 
= -\left( \frac {2\pi}V \right)^2 \left( \ekj \right)_\ell \left( \ekjp \right)_m
 \left( \ekj \right)_n \left( \ekjp \right)_p \mu^C_n \mu^C_p k k'
\nonumber \\
&\times&
\left\{ \frac 1{k+k'} \left( \frac 1{k+\k0} +\frac 1{k'+\k0} \right)
e^{i\bk \cdot \bR_{AC}} e^{i\bkp \cdot \bR_{BC}}
+  \frac {e^{i\bk \cdot \bR_{AC}} e^{-i\bkp \cdot \bR_{BC}}} 
{(k +\k0 )(k'+\k0 )} + c.c. \right\}
\label{eq:11c}
\end{eqnarray}
is the {\it nonresonant} contribution.
The nonresonant term (\ref{eq:11c}) is the same as that obtained when atom C
is in the ground state. This term has contributions from all field modes. 
The resonant term (\ref{eq:11b}), if summed over
$(\bk j, \bkp j')$, ultimately leads to the resonant term in (\ref{eq:8}): only modes
with frequency $\w0$ turn out to give a contribution to this term.
Coherently with this consideration, 
we now assume that: i) the nonresonant contribution generated by the atom C
to the field correlations  induces dipole
moments with any frequency on atoms A and B
(exactly as for three ground-state atoms);
ii) the resonant contribution induces atomic dipole moments
oscillating at frequency $\w0$ only. On this basis, in the evaluation
of the three-body Casimir-Polder potential we use a potential tensor of
the form
\begin{equation}
V_{\ell m}^{nr}(c) =
-\frac 12 F_{\ell m}^c \left( \frac {\cos kc}c + \frac {\cos k'c}c \right)
\label{eq:12}
\end{equation}
for the nonresonant field correlations, and the potential tensor
\begin{equation}
V_{\ell m}^r (c) = - F_{\ell m}^c \left( \frac {\cos \k0 c}c \right)
\label{eq:13}
\end{equation}
for the resonant field correlations, where $c=\mid \br_B -\br_A \mid$.
The latter term is specific to the case when atom C is excited and it does not appear
when the three atoms are in the ground-state. 
Thus the interaction between A and B in their ground states in the presence 
of C excited is
\begin{eqnarray}
\Delta E_{AB} &=&  \sum_{\bk j \bkp j'} 
\alpha_A(k) \alpha_B(k')
\langle E_\ell  (\bk j, \br_A) E_m  (\bkp j', \br_B) \rangle_{nr}  V_{\ell m}^{nr}(c)
\nonumber \\
&+& \sum_{\bk j \bkp j'} 
\alpha_A(k) \alpha_B(k')
\langle E_\ell  (\bk j, \br_A) E_m  (\bkp j', \br_B) \rangle_r  V_{\ell m}^r(c)
\nonumber \\
&=& -\mu_n^C \mu_p^C \alpha_A(\k0 ) \alpha_B(\k0 )
F_{\ell n}^b F_{mp}^a \frac 1{abc}
\left( \cos \k0 (a-b+c) + \cos \k0 (a-b-c) \right)
\nonumber \\
&+& \frac {\hbar c}{2\pi} F_{\ell m}^c F_{\ell n}^b F_{mp}^a \frac 1{abc}
\int_0^\infty \! du \; \alpha_A(iu) \alpha_B(iu) \alpha_C(iu)
\nonumber \\
&\times& \left( e^{-u (a+b+c)}  +\frac 14 \left( 2 + \mbox{sign}(a-c) +\mbox{sign}(b-c)
\right) e^{-u \mid a+b-c \mid} \right.
\nonumber \\
&-& \left. \frac 14 (1-\mbox{sign}(a-c)) e^{-u\mid a-b-c \mid}
- \frac 14 (1-\mbox{sign}(b-c)) e^{-u\mid a-b+c \mid} \right)
\label{eq:14}
\end{eqnarray}
Now, following the same arguments used in \cite{CP97} for ground-state atoms,
we partially symmetrize (\ref{eq:14}) over the three atoms. We must
symmetrize only the nonresonant part, for which the role of the three atoms
is indeed symmetrical; the resonant part should not be symmetrized because the 
contribution of the three atoms to this term is not symmetrical, 
only C being in an excited state
(for example, the interaction between A and C in the presence of B does not contain a 
resonant term because B is in its ground state).
Threfeore, the total three-body potential finally is
\begin{eqnarray}
\Delta E_3 &=& \frac 23 \left( \sum_{\bk j \bkp j'} 
\alpha_A(k) \alpha_B(k')
\langle E_\ell  (\bk j, \br_A) E_m  (\bkp j', \br_B) \rangle_{nr}  V_{\ell m}^{nr}(c)
+ (A \rightarrow B \rightarrow C) \right)
\nonumber \\
&+& \sum_{\bk j \bkp j'} 
\alpha_A(k) \alpha_B(k')
\langle E_\ell  (\bk j, \br_A) E_m  (\bkp j', \br_B) \rangle_r  V_{\ell m}^r(c)
\label{eq:15}
\end{eqnarray}
where $(A \rightarrow B \rightarrow C)$ indicates terms obtained from the first double sum
by permuting the atomic indices.
Explicit evaluation of (\ref{eq:15}), using (\ref{eq:14}), yields
\begin{eqnarray}
\Delta E_3 &=& -\mu_n^C \mu_p^C \alpha_A(\k0 ) \alpha_B(\k0 )
F_{\ell m}^c F_{\ell n}^b F_{mp}^a \frac 1{abc}
\left( \cos \k0 (a-b+c) + \cos \k0 (a-b-c) \right)
\nonumber \\
&+& \frac {\hbar c}\pi F_{\ell m}^c F_{\ell n}^b F_{mp}^a \frac 1{abc}
\int_0^\infty \! du \: \alpha_A(iu) \alpha_B(iu) \alpha_C(iu)
e^{-u (a+b+c)}
\label{eq:16}
\end{eqnarray}
where $a=\mid \br_B -\br_C \mid$, $b=\mid \br_C -\br_A \mid$,
$c=\mid \br_B -\br_A \mid$.
This expression is the three-atom generalization of that previously obtained
by perturbation theory for the two-atom case \cite{PT95a}.
We have obtained it with a new method, which explicitly displays the
role of dressed vacuum field correlations in the Casimir-Polder forces with excited atoms.
Compared to the case of three ground-state atoms, a resonant contribution
is now present, yielding a term in the potential oscillating in space;
this term arises from an additional contribution to the dipole correlations
of the two ground-state atoms induced by the resonant interaction of the excited
atom with the field.

In conclusion, we have shown that our model of three-body Casimir-Polder forces, 
based on dressed field correlations and
previously introduced for three ground-state atoms, can be transparently generalized 
to the case in which one of the three atoms is excited and its lifetime is long enough.
In this case we must also take into account the 
possibility of resonant processes, which give an additional (resonant) contribution to
the dipole moments induced on the atoms and to the interatomic potential.
Our method to calculate the Casimir-Polder potential, in contrast to
previous methods \cite{PT95a},
stresses the role of the spatial correlations of dressed vacuum field fluctuations,
and yields a new physical interpretation of the non-additivity of Casimir-Polder forces
for excited atoms in a time-independent framework. We plan to consider in the future
the dynamics of the system and the
consequent dynamical three-body Casimir-Polder force, by relaxing the condition we have 
imposed on the lifetime of the excited atom.

\begin{acknowledgments}
This work was supported by the European Commission under contract No. HPHA-CT-2001-40002
and in part by the bilateral Italian-Japanese project 15C1 on Quantum
Information and Computation of the Italian Ministry for Foreign Affairs. 
Partial support by Ministero dell'Universit\'{a} e della Ricerca Scientifica 
e Tecnologica and by Comitato Regionale di Ricerche Nucleari e di Struttura della Materia
is also acknowledged.
\end{acknowledgments}

\end{document}